\documentclass[12pt]{article}

\usepackage[pdftex]{graphicx}
\usepackage{subfig}
\usepackage{multicol}
\usepackage{multirow}

\def\be{\begin{equation}}
\def\ee{\end{equation}}
\def\ba{\begin{eqnarray}}
\def\ea{\end{eqnarray}}

\graphicspath{{converted_graphics/}}

\begin{document}

\baselineskip = 20pt

\title{Introducing advanced concepts for young students.}

\author{Paulo de F. Borges$^1$\thanks{pborges@cefet-rj.br}\\
\small \it $^{1}$Centro Federal de Educa\c c\~ao 
Tecnol\'ogica Celso Suckow da Fonseca\\
\small \it Coordena\c c\~ao de F\'{\i}sica, Programa de P\'os-gradua\c c\~ao em Ensino de Ci\^encias e Matem\'atica\\  
\small \it Av. Maracan\~a, 229, bloco E, quinto andar, Maracan\~a, \\
\small \it 20271-110 Rio de Janeiro, Brazil}

\maketitle

\begin{abstract}
The compound Atwood's machine (Atm) problem is revisited in order to introduce young students on advanced
concepts in Physics. Atm is an old-fashioned device. However, it allows us to speak about
relativity of motion, principle of equivalence, inertial and non-inertial frames of reference, general covariance
and invariance under coordinate transformations. Besides, it also provides experimental support for our
theoretical models. We have started with coordinate transformations and inertial and non inertial reference frames followed by the principle of equivalence.
The composed Atm was worked out in the following. We calculate the acceleration in the machine applying
Newton's Laws to describe its dynamics in an inertial frame on the fixed pulley
center and in a non-inertial one on the moving pulley center. Coordinate transformations mapping inertial frame
solutions in non-inertial ones and vice-versa allow students catch on
relativity concepts and the role played
by general covariance. A comparison between these solutions showed the importance of the
principle of equivalence in the evaluation of the intensity of gravity
locally. In addition, according to this coordinate transformation the non-inertial reference frame
equations of motion are equivalent to one single Atm plus one falling mass. Our results
have also shown, in agreement with experimental outcomes, that after movement starts single Atm
becomes lighter than $m_{3}$ mass, consequently $m_{3}$ mass is falling. It is a non-intuitive
experimental result first observed for the 1854 Poggendorff's fall machine. A measure of single Atm's mass
reduction was reported in 2016's last issue of Physics Education.
\\
\\
\noindent keywords: inertial and non-inertial frames, double Atwood Machine, equivalence principle,
coordinate transformations.

\end{abstract}

\bigskip

\section{Introduction}\label{sec1}

Physics education researchers have used the Atwood machine \cite{atwood} not only to introduce Newton's Second Law and other
related concepts, but also to investigate how students understand the involved concepts.
Students in introductory courses, who were familiar with the integral and differential calculus, had reported many difficulties with the
concepts of acceleration, external and internal forces, and the
role of the rope in the machines
after being exposed to the Atwood machine (McDermott, Shaffer and Somers, 1994).\cite{mcss}  A considerable number of
papers deal with practical uses
of the machine. Krause and Sun (2011) explain why and how strings, even without mass, exert
tension and torque over a pulley.\cite{krsu}  As the net force on the system is given by the difference
between the weights of the hanging masses, the acceleration as a function of these masses
obeys to Newtons second law. Measurements of accelerations in inertial and
non-inertial frames, using computers and PASCO Smart Pulley software, were carried out
by Chee and Hong (1999).\cite{ghho} Methods based on work and energy conservation have been used by
LoPresto (1999) to analyze the Atwood machine motion.\cite{lopr} The effects of frictional forces are discussed
by Martell and Martell (2013).\cite{martell} Variable mass systems were treated in two different ways.
Flores, Solovey and Gil (2003) used the Atwood machine to study the 
flow of sand grains through
an orifice in a container hung in the machine.\cite{fsg} Souza (2012) addressed the fall of a rope that slips
on the pulley.\cite{souza} She describes its dynamics and the Atwood machine using the Newton's second
law applied to each moving part of the system. She also applied the law to the center
of mass movement and the work-energy theorem to describe the system movement. An
Atwood machine, in which one of the bodies is allowed to swing, has deserved attention in
Physics Education and dynamical systems literature (Pujol et al 2010).\cite{pprs}
Atwood machines with
nonzero mass strings were addressed by Tarnopolski (2015).\cite{tarn} The compound Atwood machine is a
typical illustration of the use of Lagrange method (Fishbane, 1996; Fowles and Cassiday, 2004; Morin, 2008).\cite{fowday,fish,morin}
Besides,
Trumper and Gelbman (2000), have used the Atwood Machine in a microcomputer-based experiment
to demonstrate Newton's second Law with considerable precision.\cite{trge} Besides, the friction
force and the moment of inertia of the pulley can also be estimated.
West and Weliver (1999) have considered the time required for one of two identical masses
to reach the floor when the masses are released from rest, and when the variation of
earth's gravitational field with height is taken into account.\cite{wewe} They have then worked out a
frictionless rope with a uniform non zero mass per unit length hung over a pulley or circular support
to evaluate the time required for one end of the rope to reach the ground. The point of interest
was the Atwood machine scale dependence.  
Recently, an experiment
which shows the reduction of weight suffered by the machine when the bodies
hanging on the pulley are in motion, in agreement with the scale constructed
by J. C. Poggendorff\cite{pogg} in the mid-nineteenth century, has been described by Coelho, Silva and
Borges (2016).\cite{csb}
An analogy was developed between the Atwood machine, the Poggendorff device
and the Archimedes lever, in order to determine the accelerations of the
weights hanging on the pulleys. This analogy makes possible to determine
these accelerations both in the reference system of the laboratory and in an
accelerated reference frame fixed in the mobile sheave connected to the
compound Atwood machine as in the Poggendorff device. This analogy allows to determine
the relation between the dynamic variables (accelerations) and the static variables
(displacements), thus relating a situation of equilibrium with an out of this state.\cite{cbk}
Finally, in Coelho (2017) , the problem solving ways of Crawford , Gonz\'alez ,
Jafari  and Newburgh and colleagues  were combined
into a conceptual framework for the analysis of
the composite machine addressing Middle school students.\cite{coelho,craw,gonz,mateh1,mateh,newburg}
The goal is to enable these
students to acquire problem solving skills in a way that does not require heavy work with Mathematics.
For junior students, similar problems related to weak and strong principle of equivalence
were worked out in the high school by Pendrill and colleagues,\cite{pendrill}
while frames of reference were introduced in an experimental classroom activity.\cite{gross}

The subject is not exhausted in the quoted works. Advanced concepts also need to
be taught at some point in the school life of students who are interested in
careers in science and engineering. Thus, inertial and accelerated reference
frames, coordinate transformations between these systems, invariance
of physics laws, covariant formulation of physics laws, principle of equivalence,
among other important concepts that allows an
understanding of Modern Physics will be introduced through the compound Atwood machine. Initially the covariant
formulation of Mechanics is introduced in parallel with the principle of
equivalence and coordinate transformations. It is shown that Newton's Laws are
invariant under coordinate transformations between inertial reference frames and
are generally covariant when written appropriately.
Next, the classic Atwood machine is solved departing from the principle
of equivalence in a totally conceptual way. In the following session, Newton's
Laws are applied in the solution of the compound Atwood machine in the laboratory inertial
reference frame  and in an accelerated frame fixed in its
mobile sheave. The center of mass acceleration is determined by showing
that after the beginning of the movement, although the machine is in a fixed
position in space, it is globally out of mechanical equilibrium. In the
following section, the coordinate transformations that connect inertial and non-inertial
solutions are discussed with emphasis on their covariance. A
session of numerical results to compare our ones with other papers. Conclusions and
final remarks ends this manuscript.

\section{Reference Frames: Inertial and Non-inertial - The Equivalence Principle.}\label{sec2}

\subsection{Reference Frames.}\label{sub1}

In the figure bellow the vectors $\vec{r}$ and ${\vec{r}}^{\;\prime}$ locate a point $P$ in space in respect to
reference frames $O$ and $O'$ and the vector $\vec{R}$ locates $O'$ regarding $O$. These vectors are
related as follows:
\begin{equation}\label{eq1}
\vec{r}={\vec{r}}^{\;\prime}+\vec{R}\mbox{,}
\end{equation}
\begin{figure}[htb!]
\centering
\includegraphics[width=9.0cm,height=5.0cm]{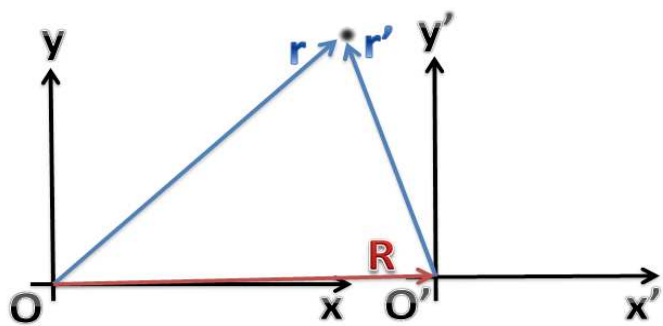}
\caption{Cartesian Reference Frames.}
\label{fig1}
\end{figure}

\noindent This relationship implies that the velocities and accelerations are given explicitly in the
following form: $\vec{v}=\vec{v'}+\vec{V}$ and $\vec{a}=\vec{a'}+\vec{A}$, after performing the time
derivatives of equation (\ref{eq1}). If $\vec{V}$ is constant, $\vec{A}=0$
and the Newton's second law is measured in both reference systems. There is no information on the
difference in dynamics observed in both systems of reference.
Besides, Galileo's transformations of
coordinates $x'=x-Vt$, $t=t'$ rise the correct results when changing from one reference frame to another.
On the other hand, when $V$ is variable and thus $A$ is non-zero, there arises a difference between
measurements made in both reference frames $O$ and $O'$ due their relative acceleration.
In this case, coordinate transformations are non-linear taking form $x'=x-(1/2)At^{2}$, $t=t'$.
We intend to use these coordinate transformations to recover correct results alternating from inertial
reference frames to non-inertial ones and reciprocally.
All inertial frames
are equivalent and match the Galilean relativity principle and all non-inertial frames that have the same
acceleration are equivalent and also match the relativity principle among them.
When reference systems have different
accelerations new terms rise into coordinate transformations which don't match the Newton's second
law.
Motion in a gravitational field under non-gravitational forces is a good example of these
situations.\cite{french,klk}
However, if there is no other reference system
to perform this comparison (coordinate transformation) or indeed, the perceptions
of an observer are based in believing it is always at rest (Aristotelian view), observers in
a non-inertial frame will explain their measures without considering any extra acceleration.
In astrology it is a matter of fact that Mars shows retrograde motion. Of course, Mars' orbit
is elliptic like the other planetary orbits. However, Earth is a non-inertial reference
frame and its acceleration should be considered when looking for Mars motion.
The coordinate transformation
from system $O$ to $O'$ introduces one term $mA$ into $P$ acceleration which is not measured neither by
$O'$ nor $O$ as a property of the body on $P$. It is a property of $O'$.
Thus, to measure centrifugal force,
the observer should be connected to the body in circular motion. An observer in
a rotating reference frame, the London Eye for example,
looking at something at rest on the ground sees an oscillatory motion.

\subsection{The Equivalence Principle and General Covariance.}\label{sub2}

Principle of equivalence is a basic principle of nature and is conceptually independent of the Newton's
second law and the law of gravitation. Its first expression manifests as the proportionality among
gravitational and inertial mass or the weak principle of equivalence:\cite{banku1}
\begin{equation}\label{eq2}
m_{i}a=\frac{Gm_{g}M_{G}}{R^{2}}\mbox{,}
\end{equation}

\noindent $m_{g}/m_{i}$ is not necessarily $1$ but $m_{g}/m_{i}=\alpha$, an universal constant. The
universality allows us to have $M_{G}=\alpha M_{i}$, consequently:
\begin{equation}\label{eq3}
a=G\alpha^{2}\frac{M_{i}}{R^{2}}=G'\frac{M_{i}}{R^{2}}\mbox{.}
\end{equation}

\noindent The experimental value of $g$, fix $G'$ value. However, Newton's second law
should be covariant under one coordinate transformation like
\begin{equation}\label{eq4}
x^{\prime}=x+\eta(t)\mbox{,}
\end{equation}

\noindent if gravity transforms as:
\begin{equation}\label{eq5}
g^ {\prime}=g+\ddot{\eta}(t)\mbox{.}
\end{equation}

\noindent According general covariance of non-relativistic laws.\cite{rosen}
In a inertial reference frame Newton's second law can be written:
\begin{equation}\label{eq6}
m_{i}\ddot{\vec{x}}_{n}=m_{g}\vec{g}+\Sigma\vec{F}(\vec{x}_{n}-\vec{x}_{j})\mbox{,}
\end{equation}

\noindent In this equation gravitational
force was considered as another external force  and $m_{i}$ is the inertial mass
and $m_{g}$ the gravitational mass.
The coordinate transformation
$\vec{x'}_{n}=\vec{x}_{n}–(1/2)\vec{A}t^{2}$, $t=t'$ belongs to equation (\ref{eq4}) class of transformations.
Submitting equation (\ref{eq6}) to this transformation and making $\vec{A}=\vec{g}$:
\begin{equation}\label{eq7}
m_{i}\ddot{\vec{x'}}_{n}=m_{g}\vec{g}-m_{i}\vec{g}+\Sigma\vec{F^{'}}(\vec{x^{'}}_{n}-\vec{x^{'}}_{j})\mbox{,}
\end{equation}

\noindent But $m_{g}/m_{i}=\alpha$ changing (\ref{eq7}) in:
\begin{equation}\label{eq8}
m_{i}\ddot{\vec{x'}}_{n}=m_{i}\vec{g}(\alpha-1)+\Sigma\vec{F^{'}}(\vec{x^{'}}_{n}-\vec{x^{'}}_{j})\mbox{.}
\end{equation}

\noindent Equation (\ref{eq8}) means the weak and strong principle of equivalence
should be considered simultaneously in order to totally eliminate the gravitational
field in the movement equation and rescuing the Newton's second law.\cite{weinberg}
We must take $(\alpha-1)=0$ or $m_{i}=m_{g}$. We then have:
\begin{equation}\label{eq9}
m_{i}\ddot{\vec{x'}}_{n}=\Sigma\vec{F^{'}}(\vec{x^{'}}_{n}-\vec{x^{'}}_{j})\mbox{.}
\end{equation}

\noindent Thus, general covariance and strong equivalence principle are connected. In short, respecting to a
free-falling frame of reference, material bodies will be non-accelerated
if they are free from non-gravitational forces. This is the same as inertial frames of reference
when there is no gravity present.\cite{mits}
Besides, globally there are no inertial reference frames. Curvature and
inhomogeneities in the gravitational field should be considered and all reference frames are
non-inertial.\cite{nar} Inertial frames should be designed locally at each point of space time in a small
region where the laws of nature take the same aspect as in no accelerated Cartesian frames in the absence of gravity.\cite{weinberg}
However, if the reference frame acceleration is not $\vec{g}$ the
gravitational field is not vanished but it is changed by this acceleration introducing a
$\vec{g}-\vec{A}$ term on the right side of eq. (\ref{eq9}), similar to equation (\ref{eq5})
preserving general covariance:
\begin{equation}\label{eq10}
m_{i}\ddot{\vec{x^{'}}}_{n}=m_{i}(\vec{g}-\vec{A})+\Sigma\vec{F^{'}}(\vec{x^{'}}_{n}-\vec{x^{'}}_{j})\mbox{.}
\end{equation}

\noindent In addition, when the observer presents acceleration other than g a
non-linear coordinate transformation works. The laws of Physics in a uniformly accelerating
system are identical to those in a inertial
system provided that we introduce a non-inertial force $F_{ni}=-mA$ in each particle.
This force is indistinguishable from the force due to a uniform
gravitational field $g=-A$. The non-inertial force, like the gravitational force, is constant
and proportional to the mass.\cite{weinberg} 
Gravitational and inertial mass are proportional and there is no way to distinguish locally
between a uniform gravitational field g and an acceleration of the coordinate system
$A=-g$. Herein locally means a sufficiently confined system.\cite{nar}
Thus, free bodies
are the ones that have no net non-gravitational forces acting on them in a frame free falling inside the field.

\section{Dynamics}\label{sec3}

In this section dynamics in inertial and non-inertial frames of reference
will be considered.

\subsection{Inertial Frame of Reference.}\label{sub3}

In our approach, acceleration is positive when it points to the ground. The string masses are much
less than the $m_{n}$ masses providing constant tension in any string point.
In addition, $m_{3}=m_{1}+m_{2}$. The tensions
over masses suspended on the moving pulley are equal, $T_{1}=T_{2}=T$, and tension in fixed
pulley string is the same tension over the mass suspended there, $T_{3}$.
\begin{figure}[htb!]
\centering
\includegraphics[width=6.0cm,height=8.0cm]{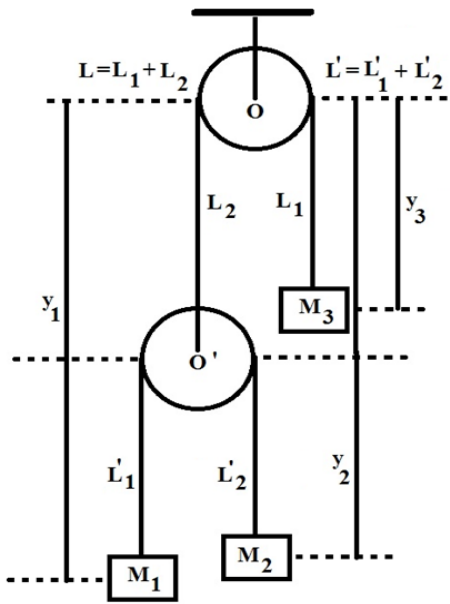}
\caption{The double Atwood machine.}
\label{fig2}
\end{figure}

\noindent Let's analyze a piece of string that at a given instant is on the moving pulley.
In this piece, string's ends act with a force equal to $T$ on the pulley. Since the string
and pulley's mass are much tinier than the other masses on the system, the sum of all forces
on the rope should tend to zero. The pulley's reaction on the sheave should be then $2T$
directed upwards. The rope passing through the fixed pulley, in turn, must apply one force
$T_{3}=2T$.\cite{saraeva} 
On the inertial frame Newton's second law and a geometrical constraint entails the following
equations of motion:
\begin{eqnarray}\label{eq11}
m_{1}a_{1}=m_{1}g-T_{1}\mbox{,}\nonumber\\
m_{2}a_{2}=m_{2}g-T_{2}\mbox{,}\nonumber\\
m_{3}a_{3}=m_{3}g-T_{3}\mbox{,}\nonumber\\
a_{1}+a_{2}=-2a_{3}\mbox{.}
\end{eqnarray}

\noindent Whose solutions are:
\begin{equation}\label{eq12}
T=\left\{\frac{4m_{1}m_{2}(m_1+m_2)}{4m_{1}m_{2}+(m_{1}+m_{2})^{2}}\right\}g\mbox{,}
\end{equation}

\begin{equation}\label{eq13}
a_1=\left\{1-\frac{4m_{2}(m_1+m_2)}{4m_{1}m_{2}+(m_{1}+m_{2})^{2}}\right\}g\mbox{,}
\end{equation}

\begin{equation}\label{eq14}
a_2=\left\{1-\frac{4m_{1}(m_1+m_2)}{4m_{1}m_{2}+(m_{1}+m_{2})^{2}}\right\}g\mbox{,}
\end{equation}

\begin{equation}\label{eq15}
a_3=-\left\{1-\frac{2(m_1+m_2)^{2}}{4m_{1}m_{2}+(m_{1}+m_{2})^{2}}\right\}g\mbox{.}
\end{equation}

\noindent Moreover, when the masses $m_{1}$ e $m_{2}$ are equal, the accelerations have ceased and $T=mg$ bringing the system to equilibrium.
The center of mass acceleration can be calculated directly. The $y$ coordinates are related by the following
expressions: $y_{1}=L_{2}+L'_{1}$, $y_{2}=L_{2}+L'_{2}$ and $y_{3}=L-L_{2}$. Keeping in mind
that $m_{1}+m_{2}=m_{3}$ and taking time derivatives algebraic manipulation results:
\begin{eqnarray}\label{eq16}
a_{CM}=\frac{1}{4}\frac{m_{1}-m_{2}}{m_{1}+m_{2}}(a_{1}-a_{2})=a_{3}\;\;\;\mbox{.}
\end{eqnarray}

\noindent Thus, the right side of equation (\ref{eq16}) is greater than zero.
The center of mass is falling and this system is globally out of equilibrium.

\subsection{Non-inertial reference frames.}
\label{sub4}

In the previous sections we have analyzed system dynamics looking from inertial frames
of reference. In this section we will repeat these calculations from a non-inertial frame of
reference. Let's start defining the reference frame and setting some properties observed
from it.
\begin{figure}[htb!]
\centering
\includegraphics[width=6.0cm,height=8.0cm]{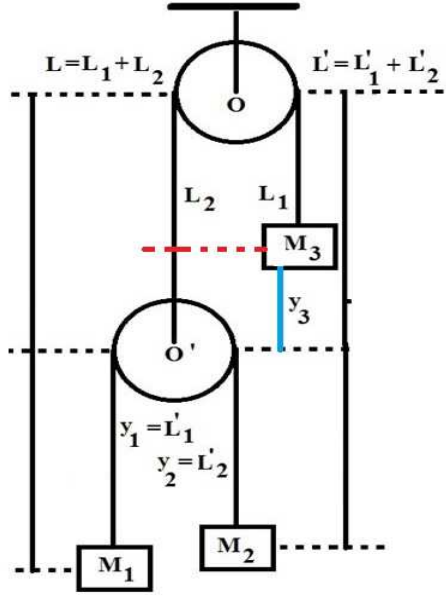}
\caption{Non-inertial view of compound Atwood machine.}
\label{fig3}
\end{figure}

\noindent The center of the moving pulley will be the place of our accelerated observer.
The observer fixed in this place is at rest, and measures mass $m_{1}$, $m_{2}$ and 
$m_{3}$ where $m_{1}+m_{2}=m_{3}$.
Assuming that for this observer  relations of cause and effect linking the
concepts of force and acceleration also exist and, moreover, that these relations can be represented
as mathematical formulae.  The observer performs measurements that will be related in a
similar way as that used by Newton.\cite{french}
For the two pulleys we have $T_{1}+T_{2}-T'=0$ and $T^{'}+T_{3}-T_{0}=0$.
If the strings are inextensible and weightless $T_{1}=T_{2}=T$ and $T^{'}=T_{3}$, the pulls
on both sides of each pulley are the same. $T^{'}$ is the force applied by rope $L$ on the
mobile pulley and $T_{0}$ the force supporting whole machine.
Kinematics is based on coordinates defined by $y_{1}=L'_{1}$, $y_{2}=L'_{2}$ and
$y_{3}=L_{1}-L_{2}$, $L=L_{1}+L_{2}$ and $L'=L'_{1}+L'_{2}$, and accelerations $\ddot{y}_{1}=-\ddot{y}_{2}=a$
and $\ddot{y}_{3}=-2\ddot{L}_{2}=a_{3}$.
The equations of motion are:
\begin{equation}\label{eq17}
m_{1}a=m_{1}g-T\mbox{,}
\end{equation}
\begin{equation}\label{eq18}
-m_{2}a=m_{2}g-T\mbox{,}
\end{equation}
\begin{equation}\label{eq19}
m_{3}a_{3}=m_{3}g-2T\mbox{,}
\end{equation}

\noindent whose solutions are:
\begin{equation}\label{eq20}
T=\frac{2m_{1}m_{2}}{m_{1}+m_{2}}g\mbox{,}
\end{equation}
\begin{equation}\label{eq21}
a=-\frac{m_{2}-m_{1}}{m_{1}+m_{2}}g\mbox{,}
\end{equation}
\begin{equation}\label{eq22}
a_{3}=\left\{\frac{m_{1}-m_{2}}{m_{1}+m_{2}}\right\}^{2}g\mbox{.}
\end{equation}

\noindent We need four equations to describe
the compound Atm for the inertial observer and three equations for the non-inertial observer.
Equations (\ref{eq20}), (\ref{eq21}) and (\ref{eq22}) seems to entail for this observer the double
machine is disassembled in a single machine and the falling $m_{3}$ mass. These problem solutions
show that for the non-inertial frame the Atwood machine motion and the $m_{3}$ motion seems independent.  
The choice of the reference frame allows us to discuss how important it is to understand motion. 
Movement is relative to the frame of reference. However, we can move from one frame to another via
coordinate transformations.
It is simpler to solve the problem using a non-inertial reference frame than an inertial frame of reference.
If masses are equal, the accelerations are
zero and $T=mg$. The system is in equilibrium. Besides, the $a_{3}$ expression is always positive
i.e. $m_{3}$ is falling in agreement with the inertial system solutions. We can include the center
of mass acceleration $a_{CM}$ in our results. Calculation is straightforward:
\begin{equation}\label{eq23}
a_{CM}=\frac{1}{2}\frac{m_{2}-m_{1}}{m_{2}+m_{1}}a_{2}+\frac{a_{3}}{2}=a_{3}\mbox{.}
\end{equation}

\noindent Again the $m_{3}$ mass is falling and this system is globally out of equilibrium.
Surprising, it is simpler
to solve the problem using a non-inertial reference frame than an inertial one.

\section{Numerical check.}\label{sec4}

In this section we will make a verification of our theoretical approach for 
a compound Atwood machine in both surveyed situations.

\subsection{Inertial values}\label{ssec1}

Considering $m_{1}=2kg$, $m_{2}=3kg$, $m_{3}=5kg$ and $g=9.8m/s^{2}$ we have got from the inertial
frame outcomes of sec. \ref{sub1} the following values for forces and accelerations: $T_{1}=T_{2}=T=24N$,
$T_{3}=2T=48N$ and $a_{1}=-2.2m/s^{2}$, $a_{2}=+1.8m/s^{2}$ and $a_{3}=+0.2m/s^{2}$,
$a_{cm}=+0.2m/s^{2}$. Besides, in this frame the moving pulley has acceleration $A=-0.2m/s^{2}$,
i.e. it is rising with the same acceleration that $m_{3}$ is falling.

\subsection{Non-inertial values}\label{ssec2}

For the non-inertial frame
sec. \ref{sub2}, the dynamics section outcomes are $T_{1}=T_{2}=T=24N$, $T^{'}=T_{3}=2T=48N$ and
$a_{1}=a=-2m/s^{2}$, $a_{2}=-a=2m/s^{2}$, $a_{3}=0.4m/s^{2}$, $a_{CM}=+0.4m/s^{2}$, when
effective gravity is $g=10m/s^{2}$.\cite{munera,gml,craw} Of course, in this situation, the fixed
pulley is seen in relative motion not only because relativity, but also because the rope is inextensible.
Its acceleration is $y_{A}=-\frac{a_{3}}{2}$ in order to maintain coherence.
These numerical results agree with Newburgh, Peidle and Rueckner and Lopes Coelho ones.\cite{newburg,lop3}
This observer measures
a gravitational field $g-A=[9.8m/s^{2}-(-0.2m/s^{2})]=10m/s^{2}$. Coordinate frames that allow
transformations in agreement with equations (\ref{eq4}) and (\ref{eq5}) are covariant.
Czudk\'ova and Musil\'ova have shown that solutions for basic problems
are simpler in non-inertial frames than in inertial ones. The observer should take account
real forces and inertial forces which preserve validity of Newton's laws.\cite{leja}
This approach could be seen as a preview of the covariance formulation of Newton's Laws.
The agreement between our numbers also reinforces
that to understand motion we have first to define the frame of reference we have in mind.

\subsection{Checking our numbers}\label{sec5}

Regarding figure (\ref{fig1}) let $\vec{A}$ be the reference frame acceleration and the
coordinate transformation $x'=x-(1/2)At^{2}$,
$t=t'$ $g^{\prime}=g-A$ from the inertial to the non-inertial one. Covariant Newton's second law is:\cite{rosen,weinberg}
\begin{equation}\label{eq24}
m_{n}\ddot{\vec{x^{'}}}=m_{n}(\vec{g}-\vec{A})+\sum{\vec{F^{'}}}(\vec{x^{'}_{n}}-\vec{x^{'}_{m}})\mbox{.}
\end{equation}

\noindent The equations of motion are transformed in the following way:
\begin{equation}\label{eq25}
 m_{1}a_{1}^{'}=m_{1}(g-A)-T_{1}\mbox{,}
\end{equation}
\begin{equation}\label{eq26}
 m_{2}a_{2}^{'}=m_{2}(g-A)-T_{2}\mbox{,}
\end{equation}
\begin{equation}\label{eq27}
 m_{3}a_{3}^{'}=m_{3}(g-A)-T_{3}\mbox{,}
 \end{equation}
\begin{equation}\label{eq28}
a_1^{'}+a_2^{'}+2a_3^{'}=-4A\mbox{.}
\end{equation}

\noindent Putting numbers, we obtain:
\begin{itemize}
\item {First for $g=g^{\prime}=9.8m/s^{2}$ the outcome is wrong. However,
it is similar to the single Atwood machine in \cite{lop3,gragra}.} 
\end{itemize}
\begin{eqnarray}\label{eq29}
2(-1.96)\neq 2(9.8+0.2)-23.52\mbox{,}\nonumber\\
3(-1.96)\neq 3(9.8+0.2)-23.52\mbox{,}\nonumber\\
5(0.392)\neq 5(9.8+0.2)-47.04\mbox{,}\nonumber\\
-1.96+1.96+2(0.392)\neq -4(-0.2)\mbox{.}
\end{eqnarray}

\begin{itemize}
\item{Here $g^{\prime}=g-A=10m/s^{2}$ our covariant transformation. The outcome is correct.
The small Atwood machine also seems disassembled from compound one.}
\end{itemize}
\begin{eqnarray}\label{eq30}
2(-2)=2(9.8+0.2)-24\mbox{,}\nonumber\\
3(-2)=3(9.8+0.2)-24\mbox{,}\nonumber\\
5(0.4)=5(9.8+0.2)-48\mbox{,}\nonumber\\
-2+2+2(0.4)=-4(-0.2)\mbox{.}
\end{eqnarray}

\noindent We can see that the sum $P_{1}+P_{2}=P_{3}$ an unexpected result since $m_{3}$ is falling.
However, the Atwood machine weight should be compared with the supporting force $2T<P_{1}+P_{2}$
on the small machine and $4T<2T+P_{3}$ on the compound one.
Furthermore, equations (\ref{eq29}) represent our check using $g=9.8m/s^{2}$ in our
non-inertial frame calculations. These values are in agreement with those calculated in section (\ref{sec2})
confirming our hypothesis that in non-inertial frames the compound Atwood machine is disassembled in
one single machine and a body freely falling. Besides, the $a_{CM}$can also be determined straightforwardly.
Globally Newton's second law is $P-4T=(m_{1}+m_{2}+m_{3})a_{CM}$ or $a_{CM}=\frac{10\times 9.8-96}{10}=0.2ms^{-2}$
for inertial frames and $a_{CM}=\frac{10\times 10-96}{10}=0.4ms^{-2}$for non inertial ones.
The following table allows us to compare our numbers:
\begin{table*}[bth]
\centering
\caption{Reference frame results.}
\begin{tabular*}{1.9\textwidth}{c c c c c}
Magnitude & \scriptsize Inertial frame & \scriptsize Non-inertial frame & \scriptsize Coelho/Graneau/Non-inertial frame & \scriptsize Newburgh et alli(2004)\\
$-$ & \scriptsize $-$ & \scriptsize $g_{e}=10m/s^{2}$ & \scriptsize $g_{e}=9.8m/s^{2}$ & \scriptsize inertial frame\\
$T$  & \scriptsize $24N$ & \scriptsize $24N$ & \scriptsize $23.52N$ & \scriptsize $T=\frac{120}{49}g=24N$\\
$a_{1}=\ddot{y}_{1}$ & \scriptsize $-2.2m/s^{2}$	& \scriptsize $+a=-2.0m/s^{2}$ & \scriptsize $+a=-1.96m/s^{2}$ & \scriptsize $a_{2}=-2.2m/s^{2}$\\
$a_{2}=\ddot{y}_{2}$ & \scriptsize $1,8m/s^{2}$	& \scriptsize $-a=2.0m/s^{2}$ & \scriptsize $-a=1.96m/s^{2}$ & \scriptsize $a_{3}=1.8m/s^{2}$\\
$a_{3}=\ddot{y}_{3}$ & \scriptsize $0.2m/s^{2}$	& \scriptsize $0.4m/s^{2}$ & \scriptsize $0.392m/s^{2}$ & \scriptsize $a_{5}=0.2m/s^{2}$\\
$a_{CM}=\ddot{y}_{CM}$ & \scriptsize $0.2m/s^{2}$ & \scriptsize $0.4m/s^{2}$ & \scriptsize $0.392m/s^{2}$ & \scriptsize $a_{CM}=0.2m/s^{2}$\\
\end{tabular*}
\label{tab1}
\end{table*}

Former works by Lopes Coelho introduce apparent weights to justify Atwood's Machine moving
out of equilibrium after their component mass motions have started.\cite{lop2}
However, center of mass acceleration is not zero and $m_{3}$ does fall in a non-intuitive way.
On the other hand, the following expression represents
the effective applied force on a single pulley machine:\cite{craw,gonz,mateh}
\begin{equation}\label{eq31}
T^{'}=2T=\frac{4m_{1}m_{2}}{m_{1}+m_{2}}g=m_{eff}g\mbox{.}
\end{equation}

\noindent 
Based on this expression they have assumed there is an effective mass:
\begin{equation}\label{eq32}
m_{eff}=\frac{4m_{1}m_{2}}{m_{1}+m_{2}}\mbox{,}
\end{equation}

\noindent where $m_{eff}$ is always less than $m_{1}+m_{2}=m_{3}$.
Making simple calculations with the table (\ref{tab1}) we obtain for $m_{eff}$ the following
value: $m_{eff}=4.8kg$. This value is less than $m_{3}=5kg$, thus $m_{3}$ must move downward.
The difference between $m_{1}+m_{2}$ and $m_{eff}$ was first shown by Poggendorff,\cite{pogg}
and measured by Graneau and Graneau and Coelho, Silva and Borges.\cite{gragra,csb}
 
\section{Conclusions.}\label{sec6}

Students have
several difficulties in learning concepts from classical and relativistic mechanics, which are
described by many researchers working in physics education research. These difficulties are related to two
critical events in the formation of the student: their previous conceptions that must be modified in the
course of the scientific practice, and the voids between pieces of technical knowledge acquired early in
their training that still does not represent a  coherent and accessible whole. When the student is able
to explain the concepts and construct a full descriptive and explanatory system based on the knowledge that
it is intended to have been learned, that is, knowledge is available to the student deliberately, and then
learning is complete.\cite{scherr,amnbor}
Previous knowledge has great influence in learning.To learn about advanced topics in science, previous
knowledge is spontaneous knowledge built in common life plus basic knowledge taught in beginners
time.\cite{ausubel, hewson}  Gauthier  suggests that powerful
principles like the equivalence principle should be introduced at the early stage of the student's
scientific development.\cite{gau} Awareness of the need to distinguish the two masses in the Newton's
second law and the law of gravitation, that equality of the two masses is not for granted and doesn't
prove Galileo's principle, several other misconceptions and knowledge learned in pieces are difficulties
that should be overcome to get successful learning.\cite{scherr}
The compound Atwood machine is a powerful tool to introduce frames of reference, principle of equivalence,
coordinate transformations and to surmount the long prevalence of flawed thinking and
the uncritical application of familiar knowledge changing the students' conceptual weaknesses. Besides, it
is an opportunity to introduce non-Newtonian points of view in Physics. We can also fill the gap among
pieces of knowledge and gain mastery on common misconceptions rose by training.\cite{scherr}

In this paper we have explored various aspects of classical mechanics, including
relativistic ones, through the device called an compound Atwood machine. The use of this device allowed discussing
important concepts such as the relativity of motion expressed in the analysis made in two different reference
systems. The concept of the reference system itself can be investigated with the introduction of accelerated
or non-inertial reference frames and its comparison with the so-called inertial reference system. The values of
the accelerations in the two systems are different, however, a transformation of variables that maps values of
values measured in one system on the other one allowed us to introduce concepts like invariance of the physics
laws and covariance of the movement equations. The validity of these transformations also allowed the introduction
of the effective gravity concept and consequently the principle of equivalence and a discussion on the
proportionality of inertial and gravitational masses. In view of these results, we believe that exploring this
simple system both in high school and in higher education is a viable alternative for teaching fundamental concepts
of mechanics in a more whole form with a great chance of success in surpassing both the barrier of previous conceptions
and the incompleteness of knowledge in pieces.
Besides, nature of science
and the role played by experiment were welcome complements to our approach.

\bigskip

\bigskip

\noindent {\bf{Acknowledgments:}}

\noindent We gratefully acknowledge that this work was supported by FAPERJ (The Foundation for Rio de Janeiro
Research Support) through grant number APV-$110.005/2013$.

\end{document}